\documentclass{rspublic}
\usepackage{psfig}

\begin{document}

\title[GRB jets dynamics]{Gamma-Ray Burst jet dynamics and their
interaction with the progenitor star}

\author[Lazzati et al.]{Davide Lazzati, Brian J. Morsony and Mitchell
C. Begelman}

\affiliation{JILA, University of Colorado, 440 UCB, Boulder, CO
80309-0440, USA}

\label{firstpage}

\maketitle

\begin{abstract}{gamma rays: bursts --  hydrodynamics -- shock waves --
supernovae: general} The association of at least some long gamma-ray
bursts with type Ic supernova explosions has been established beyond
reasonable doubt. Theoretically, the challenge is to explain the
presence of a light hyper-relativistic flow propagating through a
massive stellar core without losing those properties. We discuss the
role of the jet-star interaction in shaping the properties of the
outflow emerging on the surface of the star. We show that the nature
of the inner engine is hidden from the observer for most of the
evolution, well beyond the time of the jet breakout on the stellar
surface. The discussion is based on analytical considerations as well
as high resolution numerical simulations. Finally, the observational
consequences of the scenario are addressed in light of the present
capabilities.
\end{abstract}

\section{Introduction}

The detection of supernova signatures in the late afterglow of GRBs
(Galama et al. 1998; Stanek et al. 2003; Hjorth et al. 2003) has
established, beyond reasonable doubt, that at least a fraction of
classical long duration GRBs are associated to the final evolutionary
stages of massive stars. Such discovery has triggered speculation on
the mechanisms that allow a relativistic jet to form, propagate in the
core of a massive star, reach the surface of the star, and expand in
space without being choked by baryon entrainment.

These questions have been addressed mainly numerically, with
increasingly sophisticated codes and powerful computers. Pioneering
work (MacFadyen \& Woosley 1999) showed that a light jet can propagate
unpolluted through a massive star core by pushing the cold dense
stellar material aside. The head of the jet propagates
sub-relativistically inside the star, allowing for the shocked
material to flow in a cocoon that surrounds and hydrodynamically
collimates the outflow. These results were subsequently confirmed
by more refined special-relativistic computations (Aloy et al. 2000;
Zhang et al. 2003). Recently, Lazzati \& Begelman (2005) showed with
an analytical model that the interaction between the jet and the star
must create time evolution in the jet properties as it reach the
stellar surface, even if the energy release at the core of the star is
constant.

\begin{figure}
\psfig{file=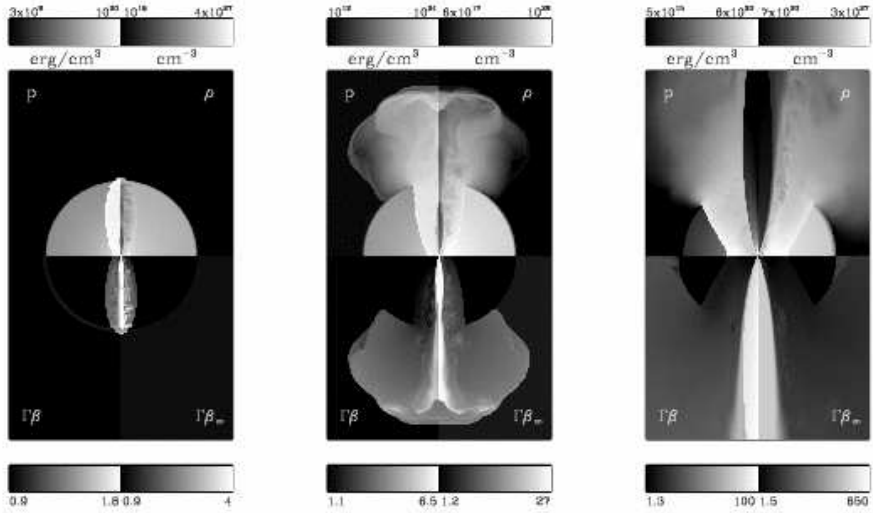,width=\textwidth}
\caption{{Stills from a simulation of a relativistic jet with
$\theta_0=10^\circ$, $L=1.33\times10^{51}$~erg~s$^{-1}$, and
$\Gamma_0=5$. The jet propagates through a star with $R=10^{11}$~cm, a
power-law density profile and a mass of $M=15M_\odot$. Each of the
three panels is divided into four sub-panels, each showing a different
quantity. Starting from the upper right panel, in clockwise order,
panels show the density, the Lorentz factor achievable at infinity, the
actual Lorentz factor and the pressure. The gray-scale is always
logarithmic. The first panel shows a still at 10.3 seconds after the
moment at which the engine is turned on. The middle panel shows a still
at 15 seconds, while the right panel shows a still at 40 seconds. The
jet is in the shocked phase in the central panel and in the un-shocked
phase in the right panel.}
\label{fig:phases}}
\end{figure}

In this paper, we show the results of high resolution two-dimensional
simulations, performed with the special relativistic Adaptive Mesh
Refinement (AMR) code FLASH. Simulations were targeted to investigate
the role of the jet-star interaction in shaping the structural and
temporal properties of the jet outside the star surface. We show that
the jet propagation takes place in four phases, three of which are
radiative and can be observed in the GRB light curve. These phases can
be separated by moments of quiescence, possibly explaining long
dead-times observed between precursors and prompt emission in GRB
light curves (Lazzati 2005). In contrast to previous lower resolution
simulations, we find that the highly relativistic part of the jet can
be well described with a sharp cutoff or with a top-hat toy model
configuration. For more details about the code, the simulations, and
some of the results reported here, see Morsony et al. (2006).

\section{The simulations}

With the aim of exploring the general properties of a confined light
jet and of understanding the features introduced by a specific stellar
progenitor we ran two different set of simulation. In one set, the
star is described as a polytropic sphere with a power-law density
profile $\rho\propto{}r^{-2.5}$. In a second set of simulations the
stellar properties are taken from model 16TI of Woosley \& Heger
(2006), which is considered a possible GRB progenitor.

The jet injection is modeled as a boundary condition on the lower edge
of the grid at $10^9$~cm. The opening angle and Lorentz factor of the
incoming jet are varied between simulations (we explore jets with
injection opening angles between 5 and $10^\circ$), but are constant
at all times in each run. The terminal Lorentz factor, or Lorentz
factor at infinity, $\gamma_\infty$, is defined as the Lorentz factor
that the material would achieve if all its internal energy were
converted to kinetic energy. It is calculated as
$\gamma_\infty=(1+4p/\rho{}c^2)\gamma$, where $\gamma$ is the local
bulk Lorentz factor. $\gamma_\infty$ for the jet material is set to
400 for all simulations and the luminosity of the central engine is
set to $1.33\times10^{51}$~erg~s$^{-1}$.  The injection Lorentz factor
is either 2 or 5. Each simulation is run for 50 seconds, giving an
energy input of $6.65\times 10^{52}$~ergs per jet, or a total energy
of $1.33\times 10^{53}$~ergs assuming symmetric jets. These total
energies are comparable to those assumed in previous works (see Zhang
et al. 2003 for a discussion).

\begin{figure}
\centerline{\psfig{file=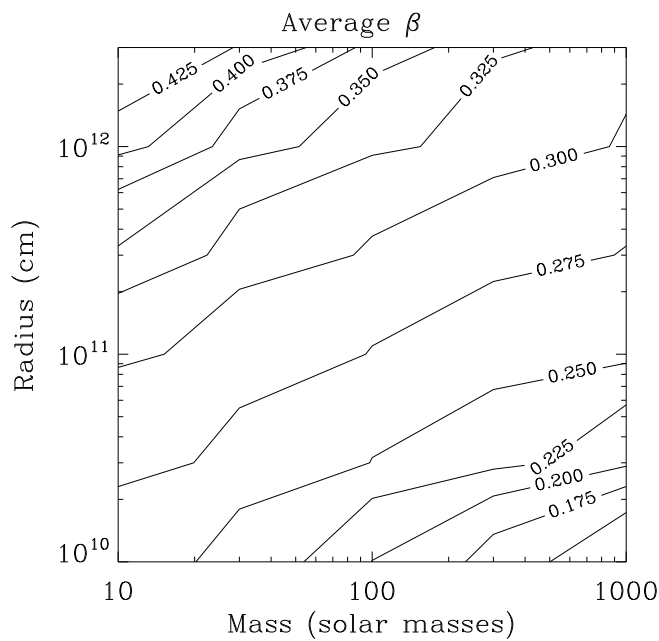,width=0.75\textwidth}}
\caption{{Average speed (in unit of the speed of light $c$) of the
propagation of the head of the jet through stars of different masses
and radii. Computations are performed with the semi-analytic model
developed by Morsony et al. (2006). Despite the large ranges in the
stellar properties, the propagation speed is always close to
$0.3c$. All jets have an initial opening angle of $10^\circ$ and an
injection Lorentz factor of $10$.}
\label{fig:betah}}
\end{figure}

\section{Results}

A visual result of simulations are shown in Fig.~\ref{fig:phases}. The
figure shows stills of the simulation for a $10^\circ$ jet propagating
in a polytropic star with $M=15M_\odot$. The figure (see also {\tt
http://rocinante.colorado.edu/$\sim$morsony/GRB/index.html} for
animations) shows how the light relativistic jet propagates through
the star and is affected by the propagation. The jet breaks on the
stellar surface at approximately 10 seconds after the injection,
releases a broad mildly relativistic component and eventually
accelerate almost to its potential Lorentz factor of $400$. We find
that the jet propagation takes place in four phases, three of which
are of great interest since are potentially radiative and can
therefore be identified in the GRB light curve.

{\bf The confined phase.}  Initially the jet is confined inside the
star. It immediately becomes supersonic and develops a shock structure
at its head. The jet head propagates subrelativistically in the star,
powered by the relativistic material behind that goes through the
reverse shock. Since the head is sub-relativistic, and well in causal
contact, the jet and stellar material do not accumulate but rather
move backward to the sides of the jet, forming a high pressure
cocoon. This cocoon pressure hydrodynamically collimates the jet
(Begelman \& Ciotti 1989). The jet propagation results therefore from
the feedback between the head working surface, that increases the
pressure of the cocoon, and the recollimation, that decreases the head
surface and weakens the cocoon (Lazzati \& Begelman 2005). Analytical
considerations for a simple jet model (Morsony et al. 2006) suggest
that the speed of the jet head is fairly independent on the stellar
properties (see also Fig.\ref{fig:betah}).

\begin{figure}
\centerline{\psfig{file=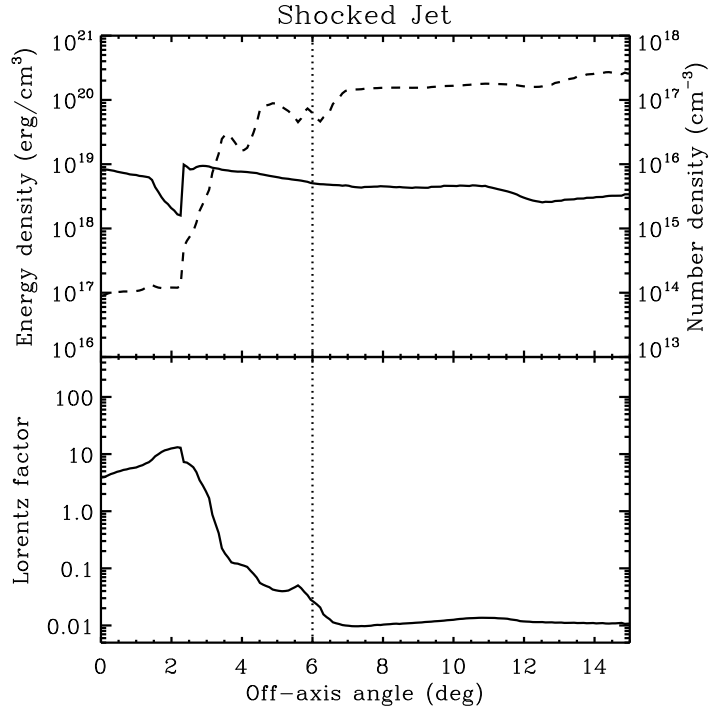,width=0.75\textwidth}}
\caption{{Cross section of the energy density and density (upper
panel) and of the quantity $\beta\Gamma$ of the shocked jet
immediately outside the stellar surface. In the upper panel, the solid
line and left y-axis show the energy density, while the dashed line
and right y-axis show the density. The jet is characterized by
variable properties and it is not easy to define a boundary between
the jet and the outside.}
\label{fig:shj}}
\end{figure}

{\bf The breakout.} As the head of the jet reaches the surface of the
star, it opens a channel through which the hot material stored in the
cocoon can escape. The cocoon material is dominated by the radiation
energy and can in principle accelerate to relativistic speed. It's
asymptotic Lorentz factor is, however, lower than that of the jet due to
mixing with the stellar baryons. Unlike the jet material, the cocoon
material is released through a nozzle on the surface of the star. As a
consequence the cocoon material forms a quasi-spherical
fireball. Taking advantage of the constancy of the propagation speed
of the jet, we can evaluate the cocoon energy to be:
\begin{equation}
E_{\rm{cocoon}} = L_j\left(t_{\rm br}-\frac{R}{c}\right) \sim 2.3 
L_j\frac{R}{c} \sim 10^{51} L_{j,50} R_{11}
\end{equation}
The energy stored in the cocoon can give rise to a precursor
(Ramirez-Ruiz et al. 2002ab). A measurement of the precursor energy
can therefore give us valuable informations on the stellar dimension,
assuming a radiative efficiency.

\begin{figure}
\centerline{\psfig{file=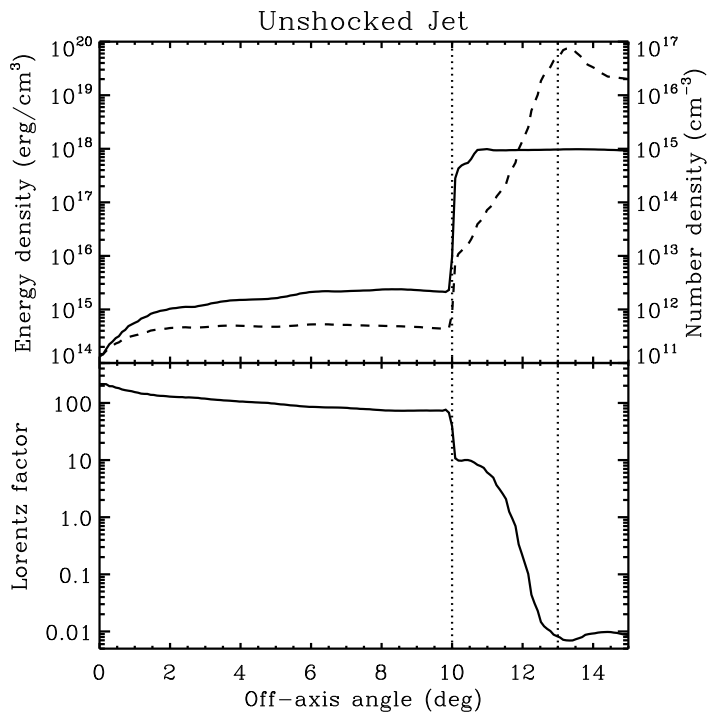,width=0.75\textwidth}}
\caption{{Same as Fig.~\ref{fig:shj} but for the un-shocked jet. In
this case, a clear structure can be identified. The core of the jet is
freely expanding with high Lorentz factor, and is separated from the
outside by a boundary layer and a shock within which the density and
speed of the material continuously approach the outside one.}
\label{fig:usj}}
\end{figure}

{\bf The shocked jet.} As mentioned above, as the jet propagates in
the confined phase, it is recollimated with tangential shocks. The
first phase of jet propagation outside the star is therefore
characterized by a jet whose structure is severely affected by the
presence of multiple shocks. Figure~\ref{fig:shj} shows a cross
section of the jet properties during the shocked phase. The cross
section is taken just outside the stellar surface, and the star model
is again the polytropic one. The jet in the figure is not clearly
defined, and the vertical dashed line is somewhat arbitrary. There are
shocks inside the jet boundary and the properties of the material are
highly variable. The opening angle during this phase is somewhat
constant, with small variations overlaid (see Morsony et al. 2006).

{\bf The un-shocked jet.} The simulations show that several tens of
seconds after the jet breaks out of the star surface, its
configuration relaxes to a much more stable one. As can be seen in
Fig.~\ref{fig:usj}, the core of the jet is made by a free streaming
outflow, accelerating according to the adiabatic expansion
$\Gamma\propto{r}$. It is unclear whether the feature in the very core
of the jet is due to the dimensionality of the computations or is
physics. This free streaming part of the jet is clearly bounded by a
strong shock. The pressure of the material outside the shock is
balanced by the jet ram pressure. The shocked area constitutes the
boundary layer between the unperturbed core and the non-relativistic
material outside. Inside the boundary layer, the density and Lorentz
factor of the material join smoothly with the outside. The width of
the layer is loosely associated to its highest Lorentz factor:
$\delta\theta\sim3^\circ\sim1/15$. The inner free streaming part of
the jet is very interesting since its propagation is unperturbed from
the inner engine out to the radiative phase. As a consequence, it
would be very interesting to isolate this component in observations
and study its variability in order to gain information on the
processes that take place at the very core of the star, where the jet
is released and energized.

{\bf The angular structure} The three phases of jet propagation
outside the star are also characterized by a different angular
structure, or energy distribution $dE/d\Omega$. During the prompt
phase any given observer receives radiation produced only from the
material moving along its line of sight. As a consequence, different
observers will see different contribution from the three phases.

\begin{figure}
\centerline{\psfig{file=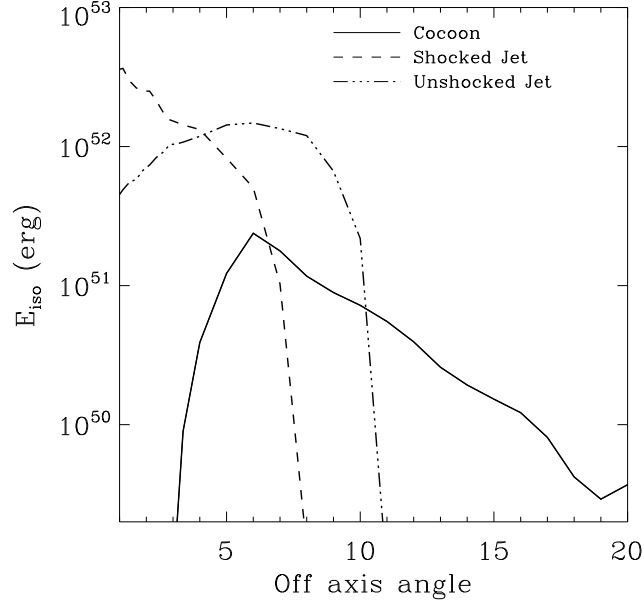,width=0.75\textwidth}}
\caption{{Energy distribution of the material ejected from the star in
the three radiative phases. The y-axis shows the isotropic equivalent
energy in ergs.}
\label{fig:dedo}}
\end{figure}

Figure~\ref{fig:dedo} shows the energy distribution in the three
phases as a function of the off-axis angle. Let us first consider an
on-axis observer. He will see a very dim precursor, if any. This is
due to the lack of energy emission in the isotropic precursor along
the jet axis. It is not entirely clear whether this is an artifact of
the 2D simulations or it is intimately linked to the process of vortex
shedding that shapes the cocoon (Morsony et al. 2006). Immediately
after the precursor, if any, the on-axis observer will enter a very
bright phase, during which the flow is dominated by the shocked
jet. This is very well collimated material, concentrated at small
angles. It is also characterized by strong variability (see
Fig.~\ref{fig:shj}) imprinted by the recollimation shocks. It should
therefore be a favorable phase for the production of $\gamma$-rays
through internal shocks and/or other dissipation mechanisms. Finally,
the on-axis observer will see emission from the un-shocked jet. This
emission is weak since most of the energy in the un-shocked phase is
stored in the boundary layer. Even though the opening angle of the jet
increases in time (Lazzati \& Begelman 2005), the luminosity of the
unshocked part of the jet is constant because the freely expanding
material is not in causal contact with the jet boundary. During this
phase no variability is imprinted in the flow by the interaction with
the star. On the other hand, this is the phase in which the intrinsic
variability of the inner engine, if any, is preserved.

Observers located at intermediate angles, with
$8^\circ<\theta<11^\circ$ in Fig.~\ref{fig:dedo}, will experience a
completely different emission pattern. They will see a relatively
bright precursor, followed by a dark phase. During the shocked jet
phase the jet is too collimated and no emission is observed at those
angles. Eventually, as the jet enters the un-shocked phase, its
opening angle spreads and the observer can see a bright GRB. The dead
time will be in this case longer than any timescale associated with
the jet or the central engine, and this effect could explain the early
precursors observed with BATSE (Lazzati 2005) and Swift.

Finally, observers located outside the maximum opening angle of the
jet (about $11^\circ$ in the case of Fig.~\ref{fig:dedo}), will only
see the precursor. Whether this weak emission could explain the
observation of faint events like GRB~980425 and GRB 031203 is however
a matter of open debate (e.g., Waxman, 2004)

\begin{figure}
\parbox{0.495\textwidth}{\psfig{file=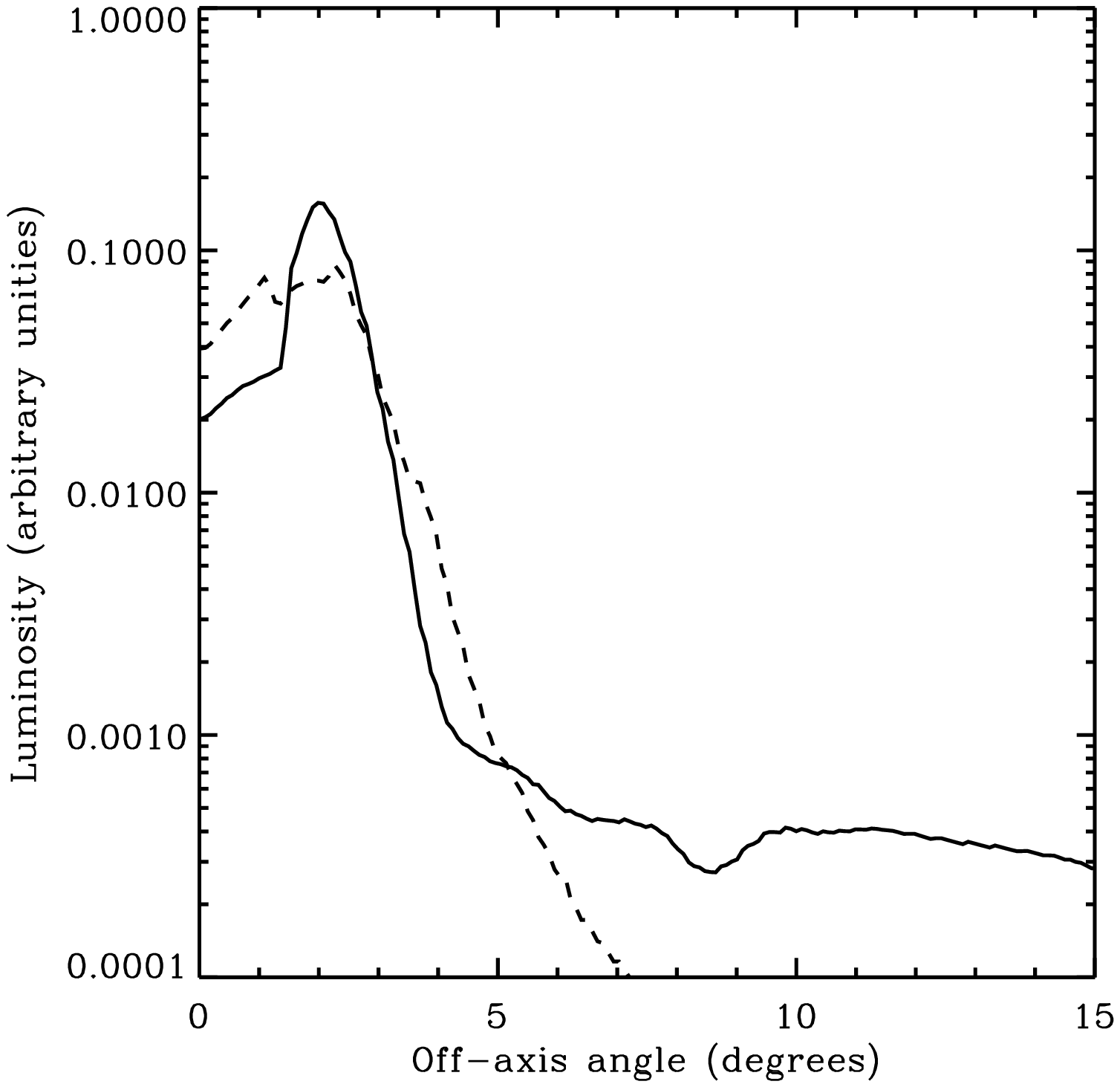,width=0.49\textwidth}}
\hspace{0.01\textwidth}
\parbox{0.495\textwidth}{\psfig{file=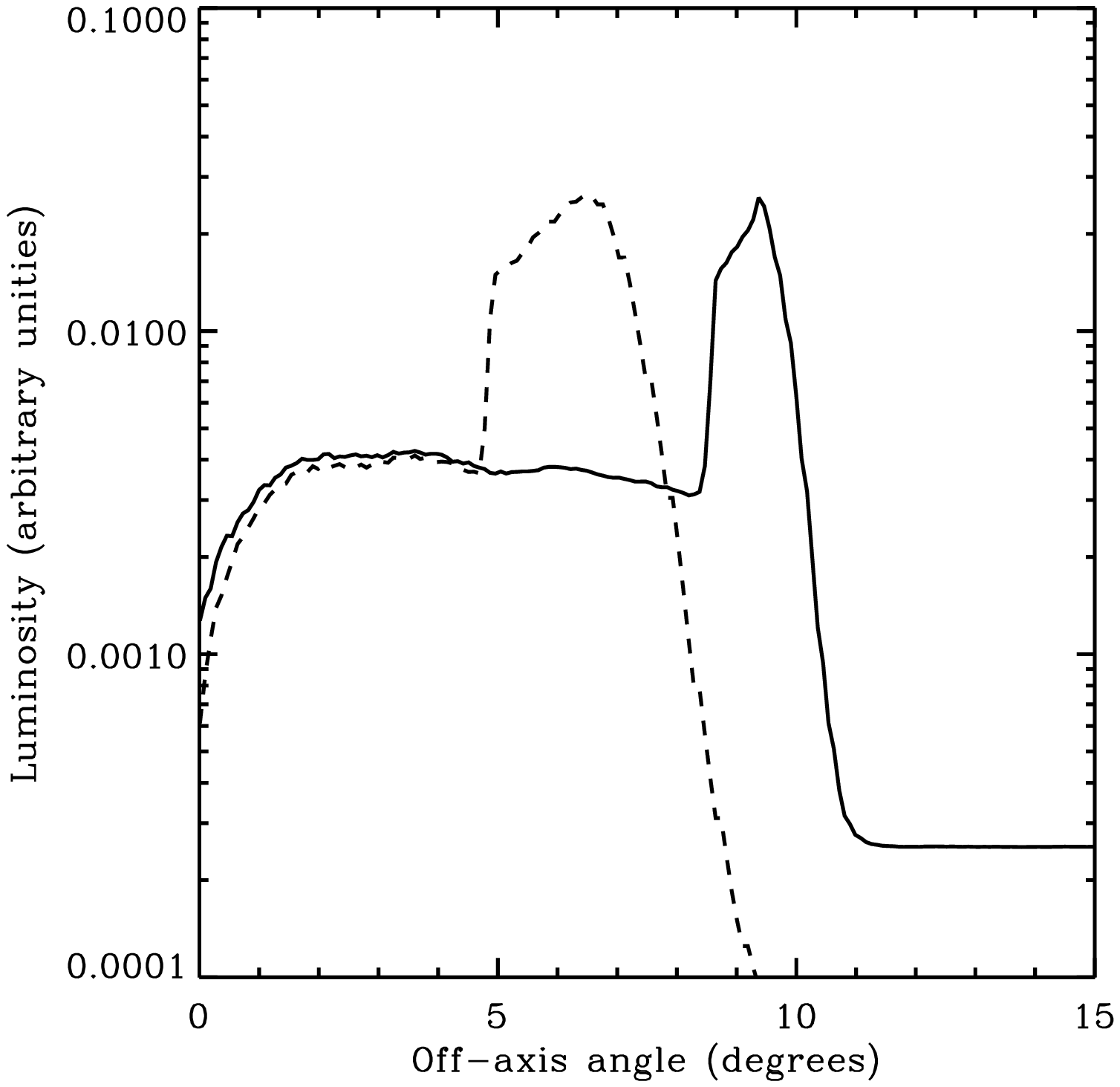,width=0.49\textwidth}}
\caption{{Evolution of the energy distribution over a factor of two in
radius for the shocked jet (left panel) and un-shocked jet phases
(right panel). Each panel shows with a solid line the energy
distribution at the star surface ($10^{11}$~cm) at time $T$. A dashed
line shows instead the distribution of the same material at
$r=2\times10^{11}$~cm at time $T+3.33$~s. The vertical axis is in
arbitrary units. Data are from the polytropic star simulations.}
\label{fig:evol}}
\end{figure}

There are two caveats to this discussion. First, the relative
normalization of the curve in Fig.~\ref{fig:dedo} can be changed
significantly. Especially, the un-shocked jet phase can last an arbitrarily
long time, and therefore can dominate most of the emission. It is
however unlikely that the inner engine will be fueled for much longer
than several hundred seconds and therefore the figure should give a
fairly correct representation of the energy budgets. Also, the
normalization of the first two phases is somewhat related to the
stellar radius and can therefore vary by approximately a factor 10.

More importantly, the energy distributions plotted in
Fig.~\ref{fig:dedo} are computed just outside the stellar radius, at
several$\times10^{11}$~cm.  The material has to travel out to the
transparency radius, located at approximately $10^{13}$~cm, before the
GRB photons are produced. To what extent the energy distribution will
be modified by the expansion over two orders of magnitude in distance
is unclear. To estimate whether there will be sizable modifications we
show in Fig.~\ref{fig:evol} the changes in the energy distribution
over a factor 2 in radius. The left panel shows the case of the
shocked jet, while the right panel shows the un-shocked jet. In both
panels we select a thin shell in the outflow and we follow it through
the expansion over a factor of two in radius, from $10^{11}$~cm to
$2\times10^{11}$~cm. The shocked jet is only marginally modified
during the expansion, the main effect being the filling of an energy
minimum in the center. The un-shocked jet, instead, changes
significantly. In particular, the strong shock that separates the
freely expanding flow from the boundary layer moves inward (in angular
coordinates), reducing the size of the freely flowing outflow. It
seems unlikely that this shrinking can proceed to completely choke the
free jet, but it is clear that energy distributions in the un-shocked
jet phase should be taken with some caution.

\section{Summary}

We presented high resolution numerical simulations of the hydrodynamic
propagation of a light relativistic jet through a massive star,
performed with the AMR code FLASH. We inject the jets as boundary
conditions at the base of the grid and follow their propagation
through the star, leaving the engine active for a timescale much
longer than the time it takes the jet to break out on the stellar
surface. We identify four phases of the jet propagation, three of
which are radiative and can possibly contribute to the GRB light curve.

We describe the temporal and angular properties of the outflow in
these three phases, showing how the outflow is initially very wide
(the precursor phase, caused by the release of the cocoon on the
stellar surface). Following this initial release, the jet emerges from
the star hydrodynamically re-collimated by tangential shocks,
producing a very bright highly collimated and variable
jet. Eventually, as the stellar influence disappears with the cocoon
release, the jet set into a stable condition whereby the core is
freely expanding and is surrounded by a boundary layer that connects
the highly relativistic core to the static dense stellar material.
The jet opening angle grows steadily in this phase. Such a scenario
allows us to explain the presence of long dead-time between the
precursor emission and the subsequent main GRB discovered by Lazzati
(2005) in approximately $20$ per cent of bright BATSE light curves.

\begin{acknowledgements}
The software used in this work was in part developed by the
DOE-supported ASC/Alliance Center for Astrophysical Thermonuclear
Flashes at the University of Chicago.  This work was supported by NSF
grant AST-0307502, NASA Astrophysical Theory Grant NNG06GI06G, and
Swift Guest Investigator Program NNX06AB69G. This work was partially
supported by the National Center for Supercomputing Applications under
grant number AST050038 and utilized the NCSA Xeon cluster.
\end{acknowledgements}

\end{document}